\documentclass[12pt]{article}
\usepackage{graphicx}
\begin{document}

\begin{center}

{\large \bf QCD. What else is needed for the Proton Structure Function?}
\vspace{5mm}

Y. S. Kim \\
Center for Fundamental Physics, University of Maryland,\\
College Park, Maryland 20742, U.S.A.
\\
E-mail: yskim@umd.edu

\end{center}

\begin{abstract}
\noindent  While QCD can provide corrections to the parton distribution
function, it cannot produce the distribution.  Where is then the starting
point for the proton structure function?  The only known source is the
quark-model wave function for the proton at rest.  The harmonic oscillator
is used for the trial wave function.  When Lorentz-boosted, this wave
function exhibits all the peculiarities of Feynman's parton picture.  The
time-separation between the quarks plays the key role in the boosting
process.  This variable is hidden in the present form of quantum mechanics,
and the failure to measure it leads to an increase in entropy.  This leads
to a picture of boiling quarks which become partons in their plasma state.
\end{abstract}

\vspace{20mm}

\noindent presented at the
Hadron Structure and QCD: from Low to High Energies \\
(Gatchina, Russia, June 30 – July 4, 2014)

\newpage

\section{Introduction}\label{intro}
While QED leads to a very successful calculation of the Lamb shift,
it cannot provide the hydrogen wave functions with their Rydberg
energy levels.  Likewise QCD gives corrections to the proton structure
function, but it cannot provide the parton distribution to which
the corrections are to be made.  The only possible starting point is
the quark-model wave function for the proton at rest.  We can then
Lorentz-boost this wave function to see whether it can provide the
starting parton distribution with all the peculiarities of Feynman's
parton picture.
\par
Do we then know how to Lorentz-boost the wave function?  This question
dates back to 1913, when Bohr started worrying about the electron orbits
in the hydrogen atom.
While Einstein was concerned with how things look to moving observers,
he did not ask the question of how those orbits would look to moving
observers.   He did not ask this question because this did not happen
in the real world.  Hydrogen atoms moving with relativistic speed did
not exist at that time.  They still do not exist.

\begin{figure}
\centerline{\includegraphics[scale=0.50]{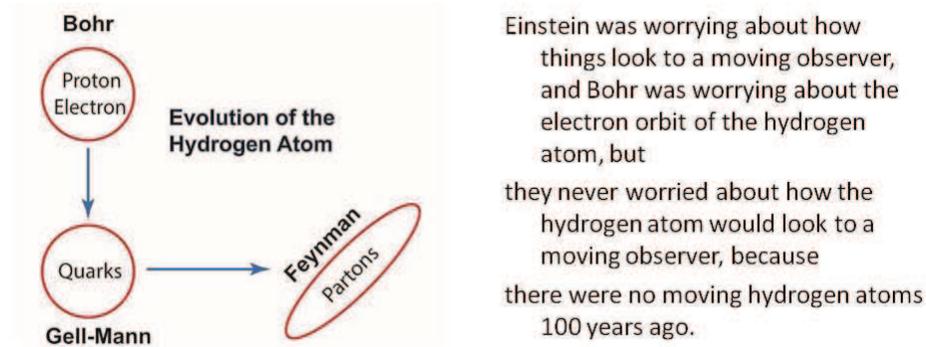}}
\caption{History of bound states in quantum mechanics.  The hydrogen
atom becomes a proton which can be accelerated to exhibits the
peculiarities of Feynman's parton picture.}\label{evol}
\end{figure}

The emergence of the quark model in 1964 changed the world.  Like the
hydrogen atom, the proton is now a bound state of more fundamental
particles called quarks.  Unlike the hydrogen atom, the proton can be
accelerated to its speed very close to that of light.  This historical
aspect is illustrated in Fig.~\ref{evol}.
\par
However, do we have wave functions that can be Lorentz-boosted?
The present form quantum field theory with Feynman diagrams has been
very successful in combining quantum mechanics with special relativity.
However, according to Feynman~\cite{fkr71}, his diagrams are not
effective in dealing with bound-state problems.  Instead, Feynman
suggested the use of harmonic oscillators.  He noted that the hadronic
mass spectra are consistent with the degeneracy of the three-dimensional
harmonic oscillator.
\par
Next question is whether it is possible to construct the oscillator
wave functions that can be Lorentz-boosted. It appears that Paul A. M.
Dirac was concerned with this question, especially in his papers of
1929, 1945, and 1949~\cite{dir27,dir45,dir49}.

\par
In Sec.~\ref{dirac}, we integrate those three papers by Dirac to
construct that the oscillator wave function that can be Lorentz-boosted.
The boosted wave function exhibits all the peculiarities in Feynman's
parton picture~\cite{fey69}.  In Sec.~\ref{feynman}, it is noted that
the time-separation variable plays the key role in the boosting process
However, it is a non-measurable hidden variable in the present form of
quantum mechanics~\cite{fey72}.  It is shown that the confined quarks
become plasma-like partons as the hadronic speed approaches that of light.

\section{Hadronic wave functions and the parton picture}\label{dirac}

In 1971, Feynman and his students noted that harmonic oscillator wave
functions with their three-dimensional degeneracy can explain the
main features of the hadronic spectra~\cite{fkr71}.  Earlier, in
1969~\cite{fey69}, Feynman proposed his parton picture where a
fast-moving hadrons appears like a collection of partons with properties
quite different from those of the quarks inside a static hadron.
\par
In their 1971 paper~\cite{fkr71}, Feynman~{\it et al.} wrote down a
Lorentz-invariant differential equation which can be separated into the
Klein-Gordon equation for a free hadron, and a harmonic-oscillator
equation for the quarks inside the hadron which determines the hadronic
mass.  Feynman's equation of 1971 contains both a running wave for the
hadron and a standing wave for the quarks inside the hadron.
\par
Their Lorentz-invariant oscillator equation takes the form
\begin{equation}\label{fkr11}
\frac{1}{2} \left[\left(\frac{\partial}{\partial x_{\mu}}\right)^2
- x_{\mu}^2 \right] \psi\left(x_{\mu}\right) =
\lambda \psi \left(x_{\mu}\right) ,
\end{equation}
where $x_{\mu}$ is the four-vector specifying the space-time
separation between the quarks.  For convenience, we ignore all
physical constants such as $c, \hbar$, as well as the spring constant
for the oscillator system.

\par
This equation has a solution of the form~\cite{dir45}
\begin{equation}\label{kn11}
\psi(z,t) =  \frac{1}{\sqrt{\pi}}
      \exp{\left\{-\frac{1}{2}\left(z^2 + t^2\right)\right\}}.
\end{equation}
This solution is Gaussian in both the $z$ and $t$ variables.  Is
it then possible to attach a physical interpretation to this wave
function.

\begin{figure}
\centerline{\includegraphics[scale=0.45]{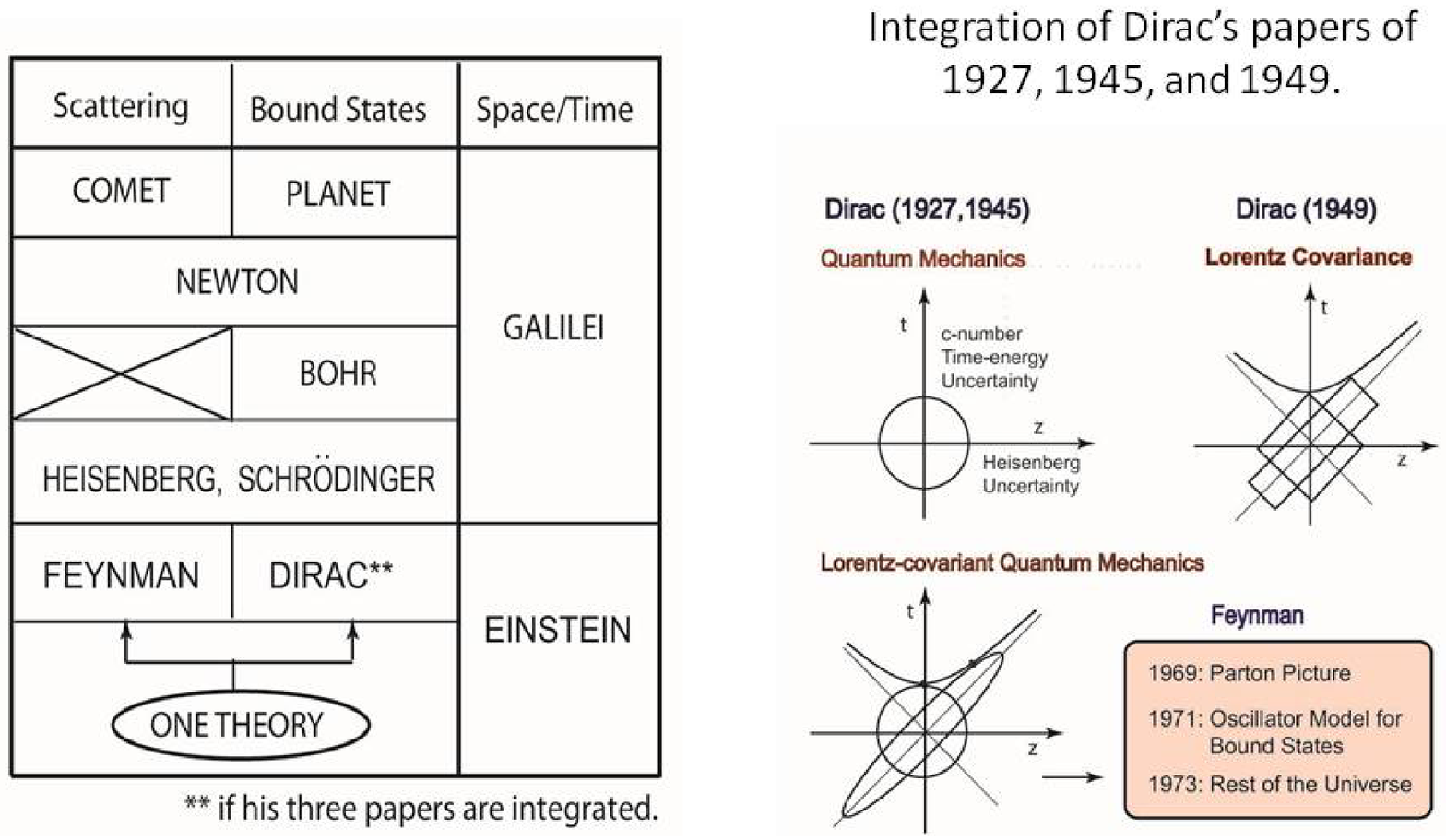}}
\caption{History of scattering and bound states.  Feynman diagrams take
care of quantum scattering problems in the Lorentz-covariant world.  For
bound-state problems, it is possible to construct harmonic oscillator wave
functions that can be Lorentz-boosted, from Dirac's ideas given in his
papers of 1927, 1945, and 1949. }\label{comet}
\end{figure}

Indeed, this Gaussian form allows us to integrate three papers
Dirac published in his attempt to construct localized wave functions
in the Lorentz-covariant world.

\begin{itemize}

\item[1.]  In his 1927 paper~\cite{dir27}, Dirac notes there is a
 time-energy uncertainty relation.  However, there are no excitations
 along the time-like axis, and it is difficult to incorporate this aspect
 to the Lorentz-covariant world.  Dirac calls this uncertainty relations
 the ``c-number'' time-energy uncertainty relation.

\item[2]   In his 1945 paper~\cite{dir45}, Dirac considers the Gaussian
  form of Eq.(\ref{kn11}) in his attempt to construct a Lorentz-covariant
  function.  However, he makes no attempts to give a physical interpretation
  to the time variable in the Gaussian expression.

  \item[3]  In his 1949 paper~\cite{dir49}, Dirac considers various forms
  of relativistic dynamics.  Among others, he says there that we can
  construct relativistic quantum mechanics by constructing a representation
  of the Poincar\'e group.  He also proposed the light-cone coordinate system.

\end{itemize}

In a series of papers since 1973~\cite{kn73}, mostly in collaboration
with my Marilyn Noz,  I was able to integrate the three papers of Dirac
listed above~\cite{knp86,kn14ps}, and the net result is summarized in
Fig.~\ref{comet}.  In order to integrate those Dirac papers, we had to
fill in the gaps among them.

\begin{itemize}
\item[1.]  For his time variable, Dirac did not mention that the $t$
  variable in his Gaussian form of Eq.(\ref{kn11}) is the time-separation
  variable. As the Bohr radius measure the distance between the proton and
  the electron, there should also be the time-like separation in the
  relativistic world.  This time separation variable is invariant under
  time translations.

\item[2.]  Then his Gaussian form can be used for his c-number uncertainty
  relation.  There still is the question of whether this c-number nature is
  consistent with relativity.  We can address this question from Wigner's
  observation that the internal space-time symmetry of massive particle in
  isomorphic to the three-dimensional rotation group~\cite{wig39}.  Thus,
  Heisenberg's position-momentum uncertainty relation can by-pass what
  happens along the time-like direction.

\item[3.]  Dirac's papers are like poems, but he never used diagrams.  It
  is possible to combine his 1927 and 1945 papers can be combined into a
  space-time distribution as specified in Fig.~\ref{comet}.  His
  Lorentz-boost in the light-cone coordinate system is also graphically
  illustrated in the same figure.  It is then straight-forward to the
  combine his quantum mechanics with relativity.
\end{itemize}

\begin{figure}
\centerline{\includegraphics[scale=0.55]{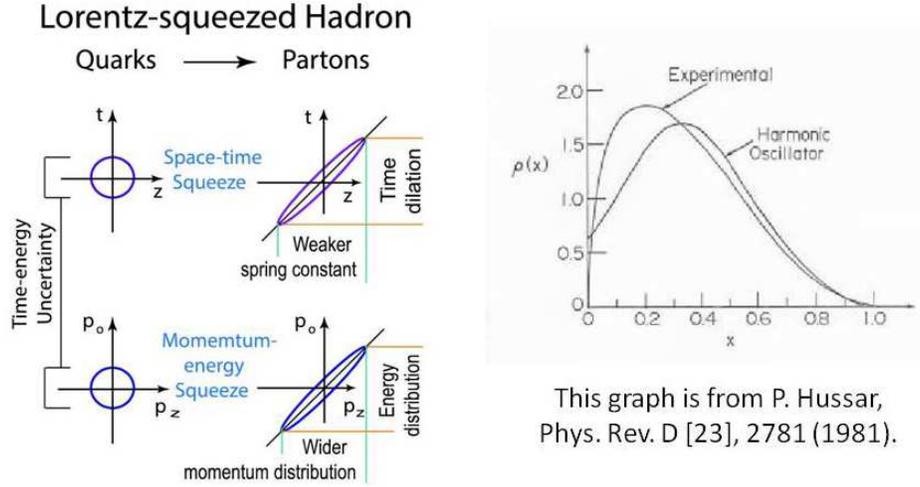}}
\caption{Lorentz-squeezed hadron and quark-parton transition
leading to a Gaussian parton distribution.  Although there
Although there is a general agreement between theory and
experiment, the disagreement is substantial.  This difference
could be corrected by QCD.}\label{parton}
\end{figure}

If the hadron moves along the $z$ direction with the velocity $v$,
the wave function of Eq.(\ref{kn11}) becomes
\begin{equation}\label{kn22}
\psi_{v}(z,t) = \frac{1}{\sqrt{\pi}}\exp{\left\{-\frac{1}{4}
   \left[\frac{c - v}{c + v}(z + t)^2
 + \frac{c + v}{c - v}(z - t)^2\right] \right\}}.
\end{equation}
This corresponds an elliptic distribution given in Fig.~\ref{comet}, where
the circular distribution is modulated by Dirac's light-cone picture of
Lorentz boosts.  The circle is ``squeezed'' into the ellipse~\cite{kn73}.

\par
According to Fig.~\ref{parton}, the quark distribution becomes concentrated
along the immediate neighborhood of one of the light cones as the hadronic
speed becomes closer to that of light. In the oscillator regime, the
momentum-energy wave function takes the same mathematical form as its
space-time counterpart.

\par
Indeed, from these Lorentz-squeezed distributions, it is possible to explain
the peculiarities of Feynman's parton picture~\cite{kn14ps,kn77par}
First,  partons are like free particles, unlike the quarks inside a hadron.
Second, the parton distribution function becomes wide-spread as the hadron
moves faster.  The width of the distribution is proportional to the hadron
momentum.  Third, the number of partons appears to be infinite.  Fourth,
the interaction time between the quarks is much longer than the interaction
time of one of the quarks with the external signal.
\par
In 1980~\cite{hwa80}, Hwa observed that the external signals do not directly
interact with the quarks, but with dressed quarks called valons.  Thus,
if we remove the valon effect, we should be able to measure the distribution
of valence quarks. With this point in mind, Hussar in 1981 compared the
parton distribution from the boosted oscillator wave function and the
experimentally measured distribution~\cite{hussar81}.  Hussar's result is
given in Fig.~\ref{parton}.
\par
As we can see in this figure, there is a general agreement between the
experimental data and the Gaussian curve derived from the Lorentz-boosted
wave function from the static quark model.  Yet, the disagreement is
substantial, especially in the small-x region, and this is the gap QCD
has to feel in.  This work is yet to be carried out.  The wave function
needs QCD to make contacts with the real world.  Likewise, QCD needs the
wave function as a starting point for calculating the parton
distribution.  They need each other.

\begin{figure}
\centerline{\includegraphics[scale=0.56]{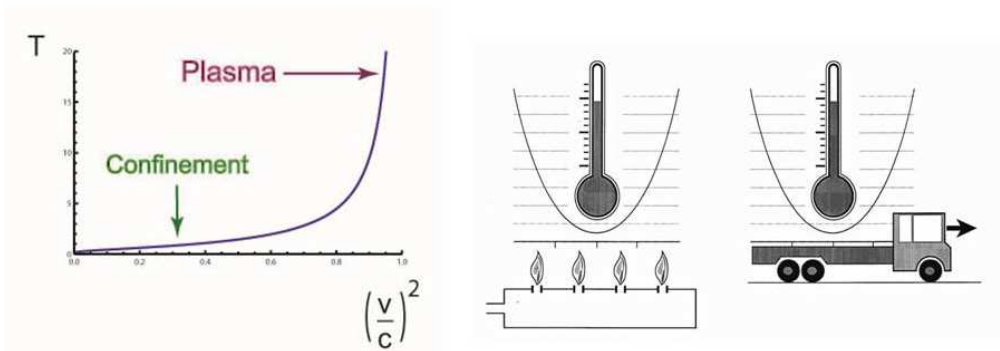}}
\caption{Lorentz boost and boiling quarks.  If the hadron is boosted
the time-like separation becomes more prominent.  If this variable is
not measured, the entropy of the system increases~\cite{kiwi90pl},
leading to a higher temperature.  At a sufficiently high temperature
the system goes through a phase transition of confined quarks to
plasma-like partons.}\label{qboil}
\end{figure}

\section{Time-separation variable and the hadronic temperature}
\label{feynman}

Let us go back to Eq.(\ref{kn11}).  The time-separation variable in the
Gaussian form is explained in terms of Dirac's c-number uncertainty relation
in Sec.~\ref{dirac}.  According to Einstein, this time separation exists
wherever there is a space-like separation like Bohr radius.  However, this
is a hidden variable not measurable in the present form of quantum mechanics.  If the variable is not measurable, we have to take a statistical
average.  In quantum mechanics, we deal with the unmeasurable variable by
talking a statistical average, or by integrating the density matrix over the
hidden variable.   We shall see in this section the consequences in the
real world of this hidden time-separation variable.

\par
The squeezed wave function of Eq.(\ref{kn22}) can be expanded
as~\cite{knp86}
\begin{equation}\label{kn33}
\psi_{v}(z,t) =  \left[1 - \left(\frac{v}{c}\right)^2\right]
    \sum \left(\frac{v}{c}\right)^n \phi_{n}(z)\phi_{n}(t) ,
 \end{equation}
where $\phi_{n}(z)$ is the oscillator wave function for its k-th excited
state. In order to deal with the time-separation variable, we construct
the density matrix
\begin{equation}
       \rho_{v}(z,t; z',t') = \psi_{v}(z,t)\psi_{v}^*(z',t') .
\end{equation}
Since we are not measuring the $t$ variable, we have to integrate over
this hidden variable:
\begin{equation}
\rho_{v}(z;z') = \int \rho_{v} (z,t;z',t) dt ,
\end{equation}
leading to
\begin{equation} \label{den77}
       \rho_{v}(z; z') = \left[1 - \left(\frac{v}{c}\right)^2\right]
    \sum \left(\frac{v}{c}\right)^{2n} \phi_{k}(z)\phi_{n}(z').
\end{equation}
This operation raises the entropy of the system~\cite{kiwi90pl}.

\par
The harmonic oscillator can also be thermally excited.  The density matrix
for the oscillator in its thermally excited state is
\begin{equation}\label{den88}
       \rho_{v}(z; z') = \left(1 - e^{-\hbar\omega/kT}\right)
    \sum e^{-n\hbar\omega/kT} \phi_{n}(z)\phi_{n}(z') .
\end{equation}
\par
Let us compare the density matrices given in Eq.(\ref{den77}) and
Eq.(\ref{den88}).  They become identical if $(v/c)^2$ is replaced by
$e^{-\hbar \omega/kT}$.  The Lorentz boost
raises the temperature of the hadron~\cite{hkn89pl}, leading to the
transition of the confined quarks to plasma-like partons~\cite{kn14ps}.
This phase transition is illustrated in Fig.~\ref{qboil}.
\par

\end{document}